\documentclass{osa-article}

\journal{oe}
\articletype{Research Article}
\begin{document}

\title{Mitigating the effect of atmospheric turbulence on orbital angular momentum-based quantum key distribution using real-time adaptive optics with phase unwrapping}

\author{Zhiwei Tao,\authormark{1,2,3} Yichong Ren,\authormark{2,4} Azezigul Abdukirim,\authormark{2} Shiwei Liu,\authormark{1,2} and Ruizhong Rao\authormark{2}}

\address{\authormark{1}School of Environmental Science and Optoelectronic Technology, University of Science and Technology of	China, Hefei 230026, China\\
\authormark{2}Key Laboratory of Atmospheric Optics, Anhui Institute of Optics and Fine Mechanics, Chinese Academy of Sciences, Hefei 230031, China\\
\authormark{3}tzw14789@mail.ustc.edu.cn\\
\authormark{4}rych@aiofm.ac.cn}

%\email{\authormark{*}rych@aiofm.ac.cn} %% email address is required



\begin{abstract}
Quantum key distribution (QKD) employed orbital angular momentum (OAM) for
high-dimensional encoding enhances the system security and information
capacity between two communication parties. However, such advantages
significantly degrade because of the fragility of OAM states in atmospheric
turbulence. Unlike previous researches, we first investigate the performance
degradation of OAM-based QKD by infinitely long phase screen (ILPS), which
offers a feasible way to study how adaptive optics (AO) dynamically corrects
the turbulence-induced aberrations in real time. Secondly, considering the
failure of AO while encountering phase cuts, we evaluate the quality
enhancement of OAM-based QKD under moderate turbulence strengths by AO after
implementing the wrapped cuts elimination. Finally, we simulate that, with
more realistic considerations, real-time AO can still mitigate the impact of
atmospheric turbulence on OAM-based QKD even in the large wind velocity
regime.
\end{abstract}

%%%%%%%%%%%%%%%%%%%%%%%%%%  body  %%%%%%%%%%%%%%%%%%%%%%%%%%
\section{Introduction}
Quantum key distribution (QKD)\cite{c1,c2,c3,c4,c5,c6} provides
unconditionally secure random numbers\cite{c7,c8,c9,c10,c11,c12} between two
authenticated distant parties (Alice and Bob) based on the no-cloning
principle of quantum physics\cite{c13,c14}. In conventional QKD schemes,
photons carrying spin angular momentum are usually used to encode information%
\cite{c15,c16,c17,c18,c19,c20}. However, the information capacity of this
binary system is limited to $1$ bit per photon, making it more vulnerable to
eavesdroppers' attacks. Unlike spin angular momentum, the orbital angular
momentum (OAM)\cite{c21} of light, associated with $l$ intertwined helical
phase $e^{il\phi }$ of each photon in a beam, where $l$ takes integer values
and is unbounded, offers an alternative way to encode information in an
infinitely large Hilbert space and improve the information capacity of QKD
system to more than $1$ bit per photon\cite{c22}. Moreover, a higher-order
system possesses a higher security threshold\cite{c23,c24,c25}, such as, for
the BB84 protocol in dimension $5$, this threshold increases to the limit of 
$21\%$ compared to $11\%$ for dimension $2$.

Over the past decade, most previous experiments, including indoors\cite%
{c26,c27,c28,c29,c30,c31,c32,c33,c34,c35} and outdoors\cite%
{c36,c37,c38,c39,c40,c41,c42}, have demonstrated such encoding schemes are
feasible up to dimension $7$\cite{c30} and the range as far as $340m$\cite%
{c38}. Nevertheless, the presence of atmospheric turbulence remains the
greatest challenge to effectively implement an OAM-based QKD experiment in
free space. Random air refractive index fluctuations caused by atmospheric
turbulence disrupt the phase alignment of original transmitted optical
fields, split the optical vortex into several individual vortices and create
the photonic OAM pairs during the propagation\cite{c43,c44,c45}, all
combinations of which result in mode scrambling at the receiver\cite%
{c46,c47,c48,c49,c50,c51,c52,c53}, making the high dimensional encoding lose
its unique advantages.

To mitigate the adverse effects of atmospheric turbulence, the common
compensation strategy is to use adaptive optics (AO) \cite%
{c38,c54,c55,c56,c57,c58}. Recently, many numerical experiments have been
performed using AO correction in quantum levels\cite{c59,c60,c61}. However,
all these scenarios are simulated by random phase screens, which is only
valid if turbulence satisfies Taylor's hypothesis\cite{c62} (i.e., when the
laser pulse width is significantly narrower than the time scale of random
refractive index fluctuations, atmospheric turbulence is assumed to be
stationary during this tiny time fraction). To overcome this barrier, we
employ the infinitely long phase screen (ILPS)\cite{c63} method to
dynamically simulate the evolution of the average quantum bit error rate and
the secret key rate of OAM-based QKD.

With the implementation of atmospheric propagation, the branch points\cite%
{c64} caused by intensity modulation will occur in the phase distribution at
the receiver, broadening the OAM spectrum distribution\cite{c45,c65} and
leading to a reduced performance of OAM-based QKD. These phase
discontinuities complicate the phase correction in an AO system. Considering
the inability of traditional AO while encountering phase cuts\cite%
{c66,c67,c68}, we evaluate the performance enhancement of OAM-based QKD by
AO after implementing the wrapped cuts elimination. We also demonstrate that
such localized manipulation is physically possible and brings a significant
improvement compared to the previously poor correction.

It should be noted that this approach brings the drawback that it reduces
the response rate of AO system\cite{c54,c69} and tends to affect the
correction performance of them while turbulence is changing rapidly\cite{c70}%
. Concretely, the slope of beacon light measured by a Hartmann wavefront
sensor\cite{c71} is not converted into the voltage signal of a deformable
mirror directly but is used to reconstruct the phase of beacon light. Then,
the reconstructed phase is repaired by phase unwrapping algorithm and
finally converted into the voltage signal of a deformable mirror. To examine
the feasibility of our scheme, we re-evaluate the performance enhancement
achieved by AO and conclude that this scheme is still able to mitigate the
impact of atmospheric turbulence on OAM-based QKD, even in the large wind
(we use the word "wind" in general referring to the crosswind) velocity
regime.

This paper is structured as follows. Sec. 2 briefly introduces the two
mutually unbiased bases (MUBs) used in OAM-based QKD and presents the
principle of OAM-based QKD that integrates the effects of atmospheric
turbulence. The main methods used in this paper, including modified ILPS,
perturbed phase correction, and wrapped cuts elimination, are detailed in
Sec. 3. In Sec. 4, we first investigate the undesirable impact of
atmospheric turbulence on a single OAM state; secondly, we evaluate the
correction effect of our enhanced AO to compensate the turbulence-induced
aberrations for OAM-based QKD under arbitrary turbulence strengths and
correction orders; finally, we consider the impact of realistic noise
contributions on the performance of OAM-based QKD. At last, Sec. 5 discusses
some deficiencies of our numerical methods and concludes this paper.

\section{Physical model}

\subsection{Photonic source}
A photon's degrees of freedom, such as time-energy, OAM, and position
momentum, are typically physical sources for implementing high-dimensional
encoding. For an OAM-based protocol, we study the two MUBs that consist of a
group of successive OAM photon states and its complementary Fourier
conjugate angular basis (i.e., so-called angular position (ANG) state) $%
\left\vert j\right\rangle =1/\sqrt{d}\sum_{l=-L}^{L}\left\vert
l\right\rangle \exp \left( -i2\pi jl/\left( 2L+1\right) \right) $, where $%
\left\vert l\right\rangle $ represents a single-photon state of a
Laguerre-Gaussian (LG) mode\cite{c72} $LG_{0,l}\left( r,\phi ,0\right) $, $%
d=2L+1$ is the dimension of the encoding subspace and $L$ is the maximum
azimuthal index used in this protocol. Without loss of generality, the LG
modes at output plane $z$ can be described in normalized cylindrical
coordinates by

\begin{eqnarray}
LG_{p,l}\left( r,\phi ,z\right) &=&\frac{A}{w_{z}}\left( \frac{\sqrt{2}r}{%
w_{z}}\right) ^{\left\vert l\right\vert }L_{p}^{\left\vert l\right\vert
}\left( \frac{2r^{2}}{w_{z}^{2}}\right)  \notag \\
&&\times \exp \left[ -\frac{r^{2}}{w_{z}^{2}}+i\left( \frac{kr^{2}}{2R_{z}}%
+l\phi -\left( 2p+\left\vert l\right\vert +1\right) \varphi _{g}\right) %
\right] ,  \label{eq1}
\end{eqnarray}%
with $A=\sqrt{2p!/\pi \left( p+\left\vert l\right\vert \right) !}$
representing the normalization constant and radial quantum number $p$ (For
more quantum properties of $p$, we refer the reader to Refs. \cite%
{c78,c79,c80,c81}), where $L_{p}^{\left\vert l\right\vert }\left( \cdot
\right) $ is the generalized Laguerre polynomial, $w_{z}=w_{0}\sqrt{1+\left(
z/z_{R}\right) ^{2}}$ is the beam waist with $w_{0}$ being the beam waist at
input plane. $z_{R}=\pi w_{0}^{2}/\lambda $ and $k=2\pi /\lambda $ denote
the Rayleigh range and the wave number respectively, $\lambda $ is the
wavelength, $R_{z}=z\left[ 1+\left( z_{R}/z\right) ^{2}\right] $ is the
radius of curvature and $\varphi _{g}$ stands for the Gouy phase associated
with propagation phase in this protocol.

\begin{figure}
	\centering
	\includegraphics[width=0.8\linewidth]{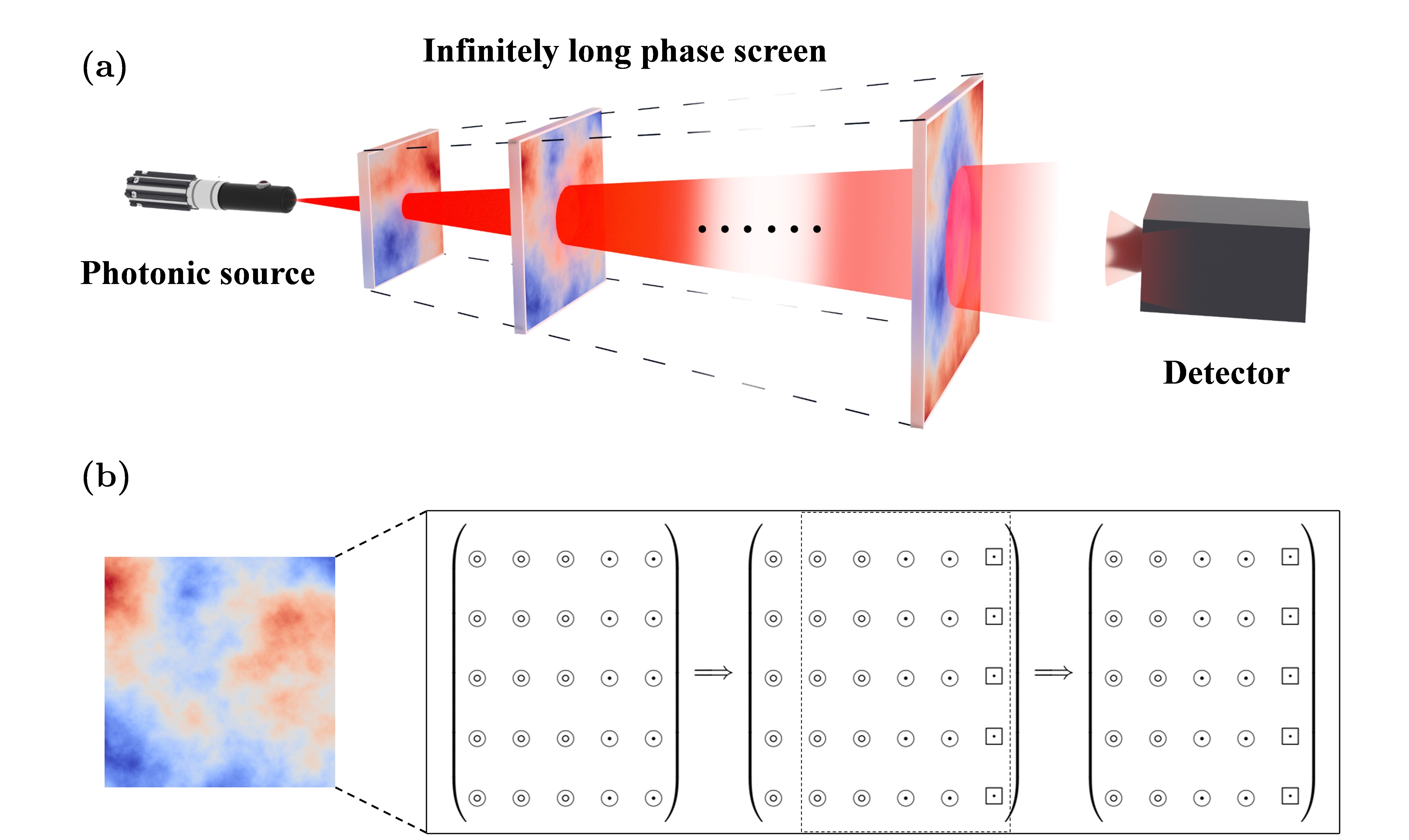}
	\caption{(a) The photonic source in Alice's laboratory randomly produces a
quantum state in either of two mutually unbiased bases and sends it to Bob
through a turbulent atmosphere (modeled by multiple layers of ILPS arranged
as ladder-shaped distribution) \ toward a detector. The phase-screen array
is ladder-shaped distributed because the divergence angle of LG modes
becomes larger with the increase of propagation distance while holding $%
\omega _{0}$ of the LG modes constant\cite{c83}. (b) A straightforward
illustration for the principle of ILPS, where the three matrices from left
to right illustrate how the last-generated phase screen connects with the
next one. The new column $X$ (illustrated by the notation $\boxdot $)
calculated from the partial data $Z$ of last-generated phase screen
(illustrated by the notation $\odot $) is added into the last-generated one
(see the second matrix). To prevent the oversized phase screen, one has to
discard the excess part that is useless (i.e., the column outside the dashed
line of the second matrix).}
\end{figure}

\subsection{Principle of OAM-based QKD}
In the weak scintillation regime, the distortion of a photonic OAM state $%
\left\vert l_{0}\right\rangle $ propagated through atmospheric turbulence
can be considered as a pure phase perturbation\cite{c46,c47}. To investigate
the influence of turbulence on OAM-based QKD under arbitrary scintillation
conditions, the split-step method\cite{c82} is used to simulate the
procedure of atmospheric propagation instead of the pure phase perturbation
approximation. As shown in Fig. 1(a), multiple layers of ILPS are arranged
consecutively to form a phase-screen array for simulating the propagation of
two MUBs across the turbulence. For convenience of presentation, the unitary
operator $\widehat{U}_{turb}^{\left( i\right) }$ is used to represent the $i$%
-th realization of a photonic state traveling through turbulence instead of
the extended Huygens-Fresnel integral\cite{c60,c61,c84}. Hence, after
undergoing the unitary transformation, the received state becomes a
superposition of several LG modes, which can be described using ket-bra
notation as follows\cite{c60,c61}%
\begin{equation}
\left\vert \psi _{l_{0}}^{\left( i\right) }\right\rangle =\widehat{U}%
_{turb}^{\left( i\right) }\left\vert l_{0}\right\rangle =\sum_{p=0}^{\infty
}\sum_{l=-\infty }^{\infty }c_{pl}^{\left( i\right) }\left\vert
pl\right\rangle ,  \label{eq2}
\end{equation}%
with $\left\vert pl\right\rangle $ corresponding to a single photon state of
a LG mode $LG_{p,l}\left( r,\phi ,z\right) $, where $c_{l,p}^{\left(
i\right) }=\left\langle pl\right\vert \widehat{U}_{turb}^{\left( i\right)
}\left\vert l_{0}\right\rangle $. Notably, since LG modes with same $w_{0}$
form an orthonormal basis, the waist of LG modes are modified to $w_{z}$\cite%
{c85} during the spectral decomposition for attributing the intermodal
crosstalk to the impact of turbulence entirely. Besides, the probability of
finding one photon in the received state with an azimuthal index $l$ can be
written as\cite{c86}%
\begin{equation}
p_{l}^{\left( i\right) }=\sum_{p=0}^{\infty }\left\vert \left\langle
pl\right\vert \widehat{U}_{turb}^{\left( i\right) }\left\vert
l_{0}\right\rangle \right\vert ^{2},  \label{eq3}
\end{equation}

Based on the principle of a single OAM state propagation, we transfer our
attention to a concrete QKD scheme. Since Alice randomly chooses her
photonic OAM states from two MUBs, we only concern the spectral broadening
of initially transmitted modes within the subspace of primary encoding basis
and regard the mode scrambling outside the subspace as the atmospheric
losses during the simulation. To get the measured matrix of OAM-based QKD,
we normalize the crosstalk probability of each incident OAM state, which can
be expressed in the coordinate representation as\cite{c73}%
\begin{equation}
p_{l_{k}\rightarrow l_{s}}^{\left( i\right) }=\frac{\sum_{p=0}^{\infty
}\int_{0}^{R}\int_{0}^{2\pi }\left\vert \psi _{l_{k}}^{\left( i\right)
}\left( r,\phi ,z\right) LG_{p,l_{s}}^{\ast }\left( r,\phi ,z\right)
\right\vert ^{2}rdrd\phi }{\sum_{l_{s}=-L}^{L}\sum_{p=0}^{\infty
}\int_{0}^{R}\int_{0}^{2\pi }\left\vert \psi _{l_{k}}^{\left( i\right)
}\left( r,\phi ,z\right) LG_{p,l_{s}}^{\ast }\left( r,\phi ,z\right)
\right\vert ^{2}rdrd\phi },  \label{eq4}
\end{equation}%
where $p_{l_{k}\rightarrow l_{s}}^{\left( i\right) }$ represents the
probability of finding OAM $l_{s}$ component from the results of projective
measurement when the incident OAM value equals to $l_{k}$, and $R$ is the
radius of receiving aperture.

In order to explore the impact of atmospheric turbulence on the performance
of OAM-based QKD, this simulation ignores the noises arising from the
detector and experimental instruments, such as dark counts\cite{c29},
afterpulsing effect\cite{c87}, to name just a few. Hence, the bit error
arises from the turbulent channel can be quantified by the average quantum
bits error rate (QBER). For the given two MUBs, the average QBER in OAM
basis is\cite{c88}%
\begin{equation}
Q_{OAM}^{\left( i\right) }=\frac{1}{d}\sum_{\substack{ l_{k},l_{s}=-L  \\ %
k\neq s}}^{L}\sum_{p=0}^{\infty }\left\vert \left\langle pl_{s}\right\vert 
\widehat{U}_{turb}^{\left( i\right) }\left\vert l_{k}\right\rangle
\right\vert ^{2},  \label{eq5}
\end{equation}%
Likewise, the average QBER in ANG basis $Q_{ANG}^{\left( i\right) }$ can
also be acquired from the same procedure. By averaging the QBER over two
MUBs, the total errors caused by atmospheric turbulence are described as 
\begin{equation}
Q^{\left( i\right) }=\frac{1}{2}\left( Q_{OAM}^{\left( i\right)
}+Q_{ANG}^{\left( i\right) }\right) ,  \label{eq6}
\end{equation}

Finally, we evaluate the information capacity that is securely exchanged
between both parties before the postselection for OAM-based QKD, quantified
by the minimum secret key rate, which can be calculated as\cite{c25,c88} 
\begin{equation}
r_{\min }^{\left( i\right) }=\log _{2}d+2\left[ Q^{\left( i\right) }\log _{2}%
\frac{Q^{\left( i\right) }}{d-1}+\left( 1-Q^{\left( i\right) }\right) \log
_{2}\left( 1-Q^{\left( i\right) }\right) \right] ,  \label{eq7}
\end{equation}

\section{Numerical methods}
\subsection{Modified infinitely long phase screen}
The simplest method to simulate atmospheric turbulence is to use the random
phase screen\cite{c89}, together with a low-spatial frequency compensation%
\cite{c90}. However, the random phase screen is established on the fact that
laser pulse width is significantly narrower than the time scale of random
refractive index fluctuations\cite{c62}. Therefore, the dynamical process of
incident light propagating through atmospheric turbulence and how AO
implements dynamical correction to the perturbation cannot be investigated
within this method (e.g., the execution rate of AO system is greater than
the variation time of turbulence). Moreover, the static assumption will lose
its usefulness when studying the influence of wind velocity on an AO-aided
OAM-based QKD.

The ILPS method gives a feasible solution to overcome this barrier. Without
loss of generality, we summarize and divide the procedure of ILPS's
execution into two steps as follows\cite{c63}: Firstly, we generate an
initially random phase screen and calculate the pixel number (i.e., rows or
columns) to be moved (we call this as the need-to-be-moved pixel number) in
each iteration according to the wind velocity $v$, wind direction $\theta $
and iteration time (i.e., time interval between two iterations). Secondly,
for the $i$-th realization of turbulence and $j$-th need-to-be-moved pixel,
we calculate the new need-to-be-added row or column $X$ from the partial
data $Z$ of last-generated phase screen, add it into the last-generated one
and discard the excess part of the oversized phase screen (a more
straightforward illustration is presented in Fig. 1(b)). The relationship
between $X$ and $Z$ is given by\cite{c63}

\begin{equation}
X=AZ+B\beta ,  \label{eq8}
\end{equation}%
where $Z$ represents the last $N_{col}$ rows/columns of last-generated phase
screen, $\beta $ is a Gaussian random vector with zero mean and its
covariance equal to unity. The matrix $A$ and $B$ (see more details for the
derivation of $A$ and $B$ and their dimension informations in Ref. \cite{c63}%
) can be calculated from the covariance of the $X$ and $Z$ vectors,
including the matrices $\left\langle ZZ^{T}\right\rangle $, $\left\langle
XZ^{T}\right\rangle $, $\left\langle ZX^{T}\right\rangle $ and $\left\langle
XX^{T}\right\rangle $. Concretly, all of covariance matrices can be acquired
from constructing the distance matrix $r_{m,n}\equiv R\left(
X_{m},Z_{n}\right) $ and acting them on the phase covariance function $%
C_{\varphi }\left( r\right) $, where $X_{m}$, $Z_{n}$ and $R\left( \cdot
\right) $ denote the $m$-th element in $X$, the $n$-th element in $Z$ and
the distance function respectively. If the turbulence spectrum follows the
von-Karman rules, $C_{\varphi }\left( r\right) $ can be expressed as\cite%
{c91}

\begin{equation}
C_{\varphi }\left( r\right) =\left( \frac{2\pi r}{L_{0}}\right) ^{\frac{5}{6}%
}\left( \frac{L_{0}}{r_{0}}\right) ^{\frac{5}{3}}\frac{\Gamma \left(
11/6\right) }{2^{5/6}\pi ^{8/3}}\left[ \frac{24}{5}\Gamma \left( \frac{6}{5}%
\right) \right] ^{\frac{5}{6}}K_{5/6}\left( \frac{2\pi r}{L_{0}}\right) ,
\label{eq9}
\end{equation}%
with the outer scale $L_{0}$ and the Fried parameter $r_{0}$, where $%
K_{5/6}\left( \cdot \right) $ is the McDonald function and $\Gamma \left(
\cdot \right) $ is the gamma function.

It is worth noting that the conventional ILPS (we use the word
"conventional" in general referring to the ILPS mentioned in Ref. \cite{c63}%
) can only simulate the effects of wind velocity changes in both horizontal
and vertical directions, which significantly limits the utilizing of this
algorithm. For this reason, we modify the conventional ILPS by decomposing
the wind velocity into horizontal and vertical directions. For simplicity,
we now detail our modifications into two parts: On the one hand, our
algorithm first detects whether a new column needs to be added in the
horizontal direction, and if so, subtracts one pixel number from the
horizontally need-to-be-moved pixel number after each motion (if we assume
the need-to-be-moved pixel number equals to $N_{move}$, then the
horizontally and vertically need-to-be-moved pixel number equal to $%
N_{move}\sin \theta $ and $N_{move}\cos \theta $ respectively), then,
executes the same procedure in the vertical direction and repeats the two
steps until no motion demand is required in either direction. On the other
hand, after undergoing above manipulation, the residual pixel number of two
directions (i.e., the difference between the horizontally/vertically
need-to-be-moved pixel number and the physically moved pixel number in two
directions), which is commonly less than one pixel number, is added to the
horizontally and vertically need-to-be-moved pixel number for the next
iteration\cite{c92} (see two examples of modified ILPS in Fig. 2).

\begin{figure}
	\centering
	\includegraphics[width=0.8\linewidth]{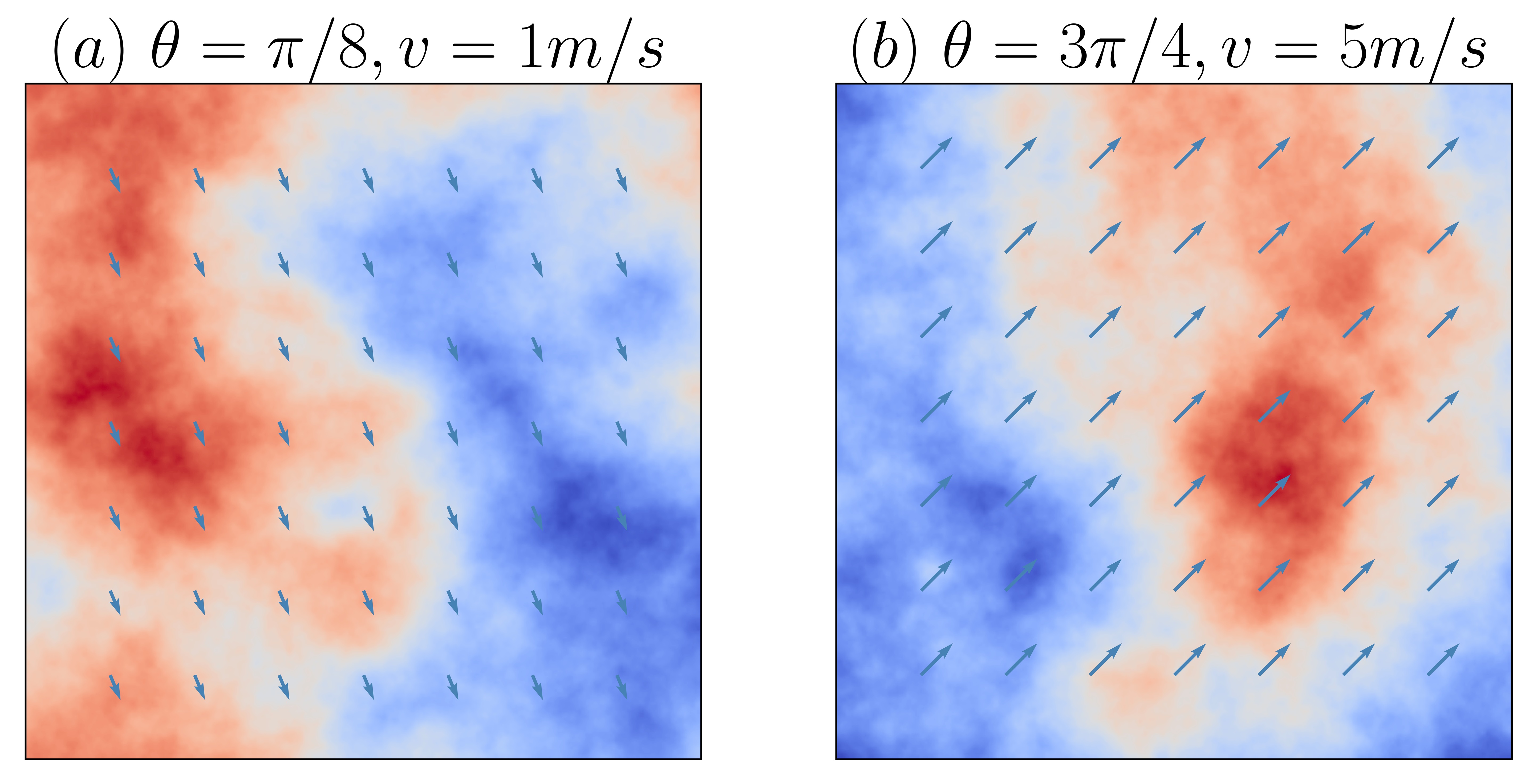}
	\caption{Two independent examples of modified ILPS with different wind
directions and velocities (a) $\theta =\pi /8$ and $v=1m/s$ (see
Visualization 1) (b) $\theta =3\pi /4$ and $v=5m/s$ (see Visualization 2).
The direction and the length of the arrows represent the wind direction and
velocity respectively.}
\end{figure}

\subsection{Adaptive optics system}

\subsubsection{Perturbed phase correction}

A well-developed AO system comprises three primary assignments: wavefront
measurement, reconstruction, and beam stabilization. The slope magnitude of
beacon light is measured by a Hartmann wavefront sensor, and subsequently
used to invert the voltage signal, which is used to generate the phase
distribution of a deformable mirror, through a reconstructor. For a
closed-loop AO system, the estimated phase is also fed back to compensate
the slope magnitude\cite{c93}, the combinations of which can continuously
improve the phase reconstruction accuracy.

Since the phase generated by AO system is only an approximation of the
perturbed phase of beacon light, the correction capability of entire system
can be commonly measured by the number of coefficient that is calculated
from the spectral decomposition of beacon light using Zernike polynomial\cite%
{c57,c58}. (e.g., for an ideal AO system, the correction order can reach
infinity because the perturbed phase of beacon light can be expanded to a
superposition of an infinite number of Zernike polynomial with different
orders, which is usually considered as an upper boundary for the performance
enhancement of AO). For a realistic AO system with $N$-th order correction
capability, the phase estimated by a deformable mirror can be expressed as%
\begin{equation}
\widetilde{\varphi }^{\left( i\right) }\left( r,\phi ,z\right)
=\sum_{n=1}^{N}a_{n}^{\left( i\right) }\left( z\right) Z_{n}\left( \frac{r}{R%
},\phi \right) ,  \label{eq10}
\end{equation}%
where $\widetilde{\varphi }^{\left( i\right) }\left( r,\phi ,z\right) $ is
the estimated phase of beacon light for the $i$-th realization of turbulence
and $Z_{n}\left( \cdot \right) $ is the Zernike polynomial with $n$-th
order. The coefficients $a_{n}^{\left( i\right) }\left( z\right) $ are given
by the overlap integral

\begin{equation}
a_{n}^{\left( i\right) }\left( z\right) =\int_{0}^{R}\int_{0}^{2\pi }\varphi
^{\left( i\right) }\left( r,\phi ,z\right) Z_{n}\left( \frac{r}{R},\phi
\right) rdrd\phi ,  \label{eq11}
\end{equation}%
where $\varphi ^{\left( i\right) }\left( r,\phi ,z\right) $ represents the
perturbed phase of beacon light. After AO correction, the estimated phase
will imprint on the single OAM photon state. We express the action of AO by
unitary operator $\widehat{U}_{AO}^{\left( i\right) }\equiv \exp \left\{ -i%
\widetilde{\varphi }^{\left( i\right) }\left( r,\phi ,z\right) \right\} $
with ket-bra notation\cite{c60,c61} 
\begin{equation}
\left\vert \widetilde{\psi }_{l_{0}}^{\left( i\right) }\right\rangle =%
\widehat{U}_{AO}^{\left( i\right) }\widehat{U}_{turb}^{\left( i\right)
}\left\vert l_{0}\right\rangle ,  \label{eq12}
\end{equation}

Finally, we emphasize that, in our AO-aided OAM-based QKD scheme, the signal
part of incident modes consists of photonic OAM states and their Fourier
conjugate angular states (i.e., ANG states). Besides, we employ the platform
beam as a probe state to detect atmospheric turbulence\cite{c94}. Our
simulations ensure that two beams are emitted simultaneously and propagated
coaxially to provide a better correction for the signal part.

\subsubsection{Wrapped cuts elimination}

A realistic AO system has been proven to be ineffective at compensating for
branch points that occur at places of zero amplitude in an optical field\cite%
{c95}. These unavoidable phase cuts\cite{c64} degrade the performance of
OAM-based QKD due to the inability of a continuous surface\ deformable
mirror to reconstruct the discontinuous phase\cite{c66,c67,c68}. Besides,
some previously correlated researches have demonstrated that branch points
begin to appear when the Rytov variance exceeds $0.1$\cite{c96,c97}.
Reassuringly, this adverse effect is not completely insurmountable. Inspired
by Refs. \cite{c98}, we employ the phase unwrapping algorithm proposed in
Ref. \cite{c99} to enhance the performance of AO correction in this paper.

The main idea of our enhanced AO scheme is that we reconstruct the perturbed
phase using the real-time slope information of beacon light and then unwrap
the discontinuous phase by an effective and implementable algorithm;
subsequently, the unwrapped perturbed phase is used to generate the voltage
signal of a deformable mirror. Generally, a standard unwrapping operation on
the perturbed phase $\varphi ^{\left( i\right) }\left( r,\phi ,z\right) $
can be accomplished by adding an integer multiple of $2\pi $ at each pixel
of wrapped phase, which is mathematically expressed as\cite{c100}

\begin{equation}
\Phi ^{\left( i\right) }\left( r,\phi ,z\right) =\left\{ 
\begin{tabular}{cc}
$\varphi ^{\left( i\right) }\left( r,\phi ,z\right) +2\pi k\left( r,\phi
,z\right) $ & $\left( r,\phi \right) \in G$ \\ 
$\varphi ^{\left( i\right) }\left( r,\phi ,z\right) $ & $\left( r,\phi
\right) \notin G$%
\end{tabular}%
\right. ,  \label{eq13}
\end{equation}%
where $\Phi ^{\left( i\right) }\left( r,\phi ,z\right) $ is the unwrapped
perturbed phase, and $k\left( r,\phi ,z\right) $ is the integer that needs
to be solved, regarded as a multi-class classification problem, $G$ stands
for the set of minimum neighborhood containing the wrapped cuts. Considering
the perturbed phase varies with a period of $2\pi $, we indicate that the
above operation is physically feasible. After implementing phase unwrapping,
the decomposition coefficient is modified and can be re-evaluated by

\begin{equation}
A_{n}^{\left( i\right) }\left( z\right) =\int_{0}^{R}\int_{0}^{2\pi }\Phi
^{\left( i\right) }\left( r,\phi ,z\right) Z_{n}\left( \frac{r}{R},\phi
\right) rdrd\phi ,  \label{eq14}
\end{equation}%
which leads to a significant performance enhancement compared to the
previously poor correction.

\begin{figure}
	\centering
	\includegraphics[width=\linewidth]{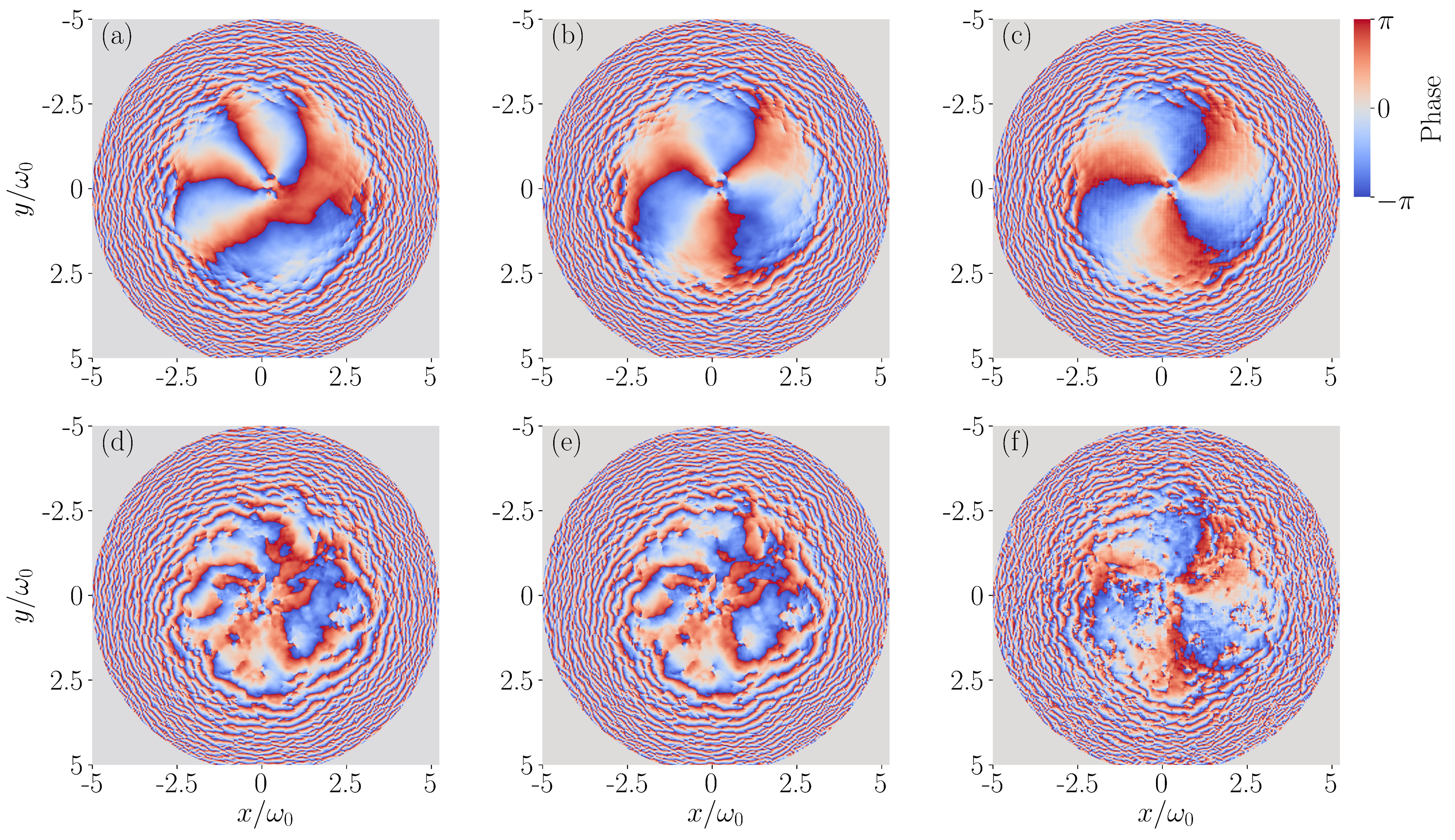}
	\caption{Phase distribitions for a single OAM state with azimuthal index $l=3
$, for a single atmospheric propagation without correction ((a) and (d)),
with realistic correction ((b) and (e)) and ideal correction ((c) and (f)),
the upper and lower three plots represent the realization in the weak and
strong scintillation regime corresponding to $\sigma _{R}^{2}=0.153$ and $%
2.235$, respectively, where $\sigma _{R}^{2}$ represents the scintillation
strength, quantified by the Rytov variance. Phase distributions repaired by
our enhanced AO with (b) $30$-order correction and (e) $50$-order
correction. The colorbar for all phase distribution plots (a)--(f) is the
same.}
\end{figure}

\section{Results}

\subsection{A single OAM state propagation}
To quantify the undesirable impact of atmospheric turbulence on OAM-based
QKD and what effects can be obtained through partial correction by AO, the
evolution of a single OAM state that is perturbed while traveling across the
turbulent channel and repaired by a specific-order AO are investigated. In
Fig. 3, we plot the phase distributions of a state $l_{0}=3$ for a single
realization of turbulence with no correction (Fig. 3(a) and 3(d)), realistic
correction\cite{c101} (Fig. 3(b) and 3(e)) and ideal correction (Fig. 3(c)
and 3(f)), where the upper and lower three diagrams represent the results
realized in the weak and strong scintillation regime, respectively (see more
details about the parameter settings in Ref. \cite{c102}).

Fig. 3(a) shows that the phase distribution of OAM state is slightly
distorted after propagation. However, this circumstance becomes more severe
in the strong scintillation regime (Fig. 3(d)), which will lead to a
completely disrupted structure. Notably, we observe in Fig. 3(a) that the
initial vortex splits into three individual vortices, accompanied by the
vortex-antivortex pairs regeneration at places of central point of initial
one. Besides, turbulence also leads to a longer discontinuous cut in the
perturbed phase distribution (For more detailed explanations, we refer the
reader to Refs. \cite{c64,c65}). To eliminate the phase discontinuities, a
realistic AO is employed to repair the perturbed phase distribution, as
shown in Fig. 3(b) and 3(e) ((b) $30$-order realization (e) $50$-order
realization). It is highlighted that, in the weak scintillation regime, the
longest phase cut that occurs in the perturbed phase is completely
eliminated, making the unwrapped corrected phase is more approach to the
unperturbed one (see Fig. 3(b)). On the contrary, in the strong
scintillation regime, the corrected phase remains almost unchanged compared
to the perturbed one (see Fig. 3(e)). Fortunately, such circumstance is
significantly alleviated in ideal correction (see Fig. 3(f)).

The poor correction is likely because the length of branch cuts becomes a
crucial factor in affecting AO correction effectiveness. In other words, in
the weak scintillation regime, only a few branch cuts are generated such
that these can be ignored when we implement realistic AO correction (see
more discussions in Ref. \cite{c45}). However, although we eliminate all
wrapped cuts in the strong scintillation regime, the accumulation of branch
cuts will destroy the phase distribution of the beacon light eventually.

In order to evaluate the correction effect of AO intuitively, the crosstalk
probability distributions (including the radial-mode scrambling) averaged
over $500$ realizations of turbulence are presented in Fig. 4. We observe
that, in the weak scintillation regime, our enhanced AO scheme effectively
restores the power into the initial OAM state even in the presence of phase
cuts. Conversely, in the strong scintillation regime, we note that our
enhanced AO loses its advantage to mitigate turbulence-induced crosstalk
with the increase of branch points. Furthermore, we also find the well-known
result\cite{c61,c103,c104} that the spectral broadening peaks around the
incident state. In the strong scintillation regime, the crosstalk
distribution has two peaks around $l_{0}=3$ and $l_{0}=-3$ (For more
discussions about the bimodal distribution, we refer the reader to Ref. \cite%
{c61}), which results in a smaller average OAM of the received state
compared to that of the incident one\cite{c43}.

\begin{figure}
	\centering
	\includegraphics[width=\linewidth]{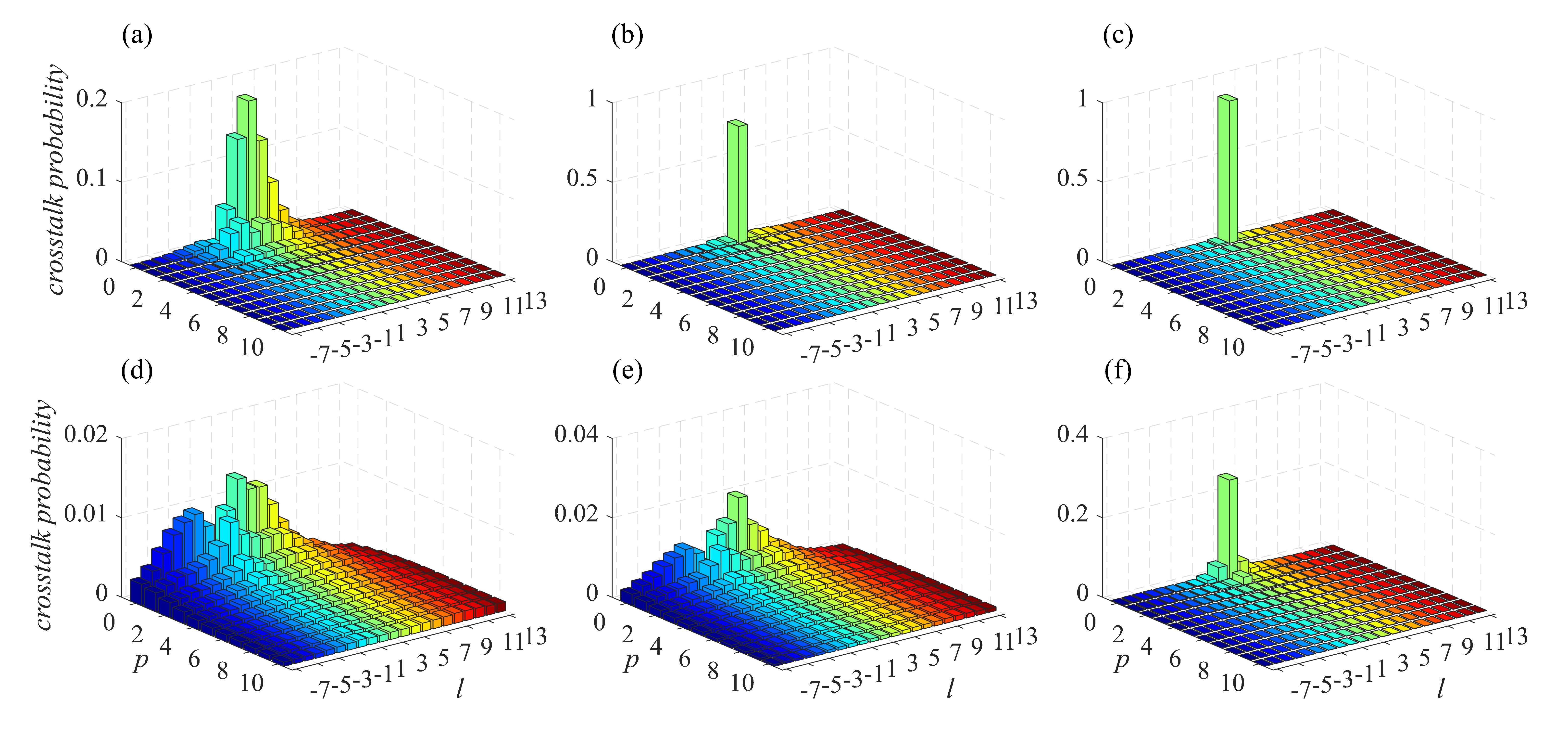}
	\caption{Crosstalk probability distributions, including radial-mode
scrambling, averaged over $500$ realizations of turbulence with azimuthal
index $l=3$, for different degrees of correction (a), (d) no AO; (b), (e)
realistic AO; (c), (f) ideal AO, the upper and lower three plots represent
the realization in the weak and strong scintillation regime, respectively.
All parameter settings are same as in Fig. 3.}
\end{figure}

\subsection{Performance of OAM-based QKD across turbulence}
Based on the above analysis for the propagation characteristics of a single
OAM state, we now employ our enhanced AO to improve the quality of OAM-based
QKD (without (a) and with (b) increasing the mode spacing in encoding
subspace). Fig. 5 illustrates the variation curves of average QBER without,
with realistic and with ideal AO correction recorded during half an hour for
5-dimensional OAM-based QKD system. we set the correction capability of
realistic AO to $30$-order and $C_{n}^{2}$ to $2.2\times 10^{-15}m^{-2/3}$
during this simulation, where $C_{n}^{2}$ is the refractive index structure
constant (see more detailed parameter settings in Ref. \cite{c102} and
reasons in Ref. \cite{c105}).

\begin{figure}
	\centering
	\includegraphics[width=0.85\linewidth]{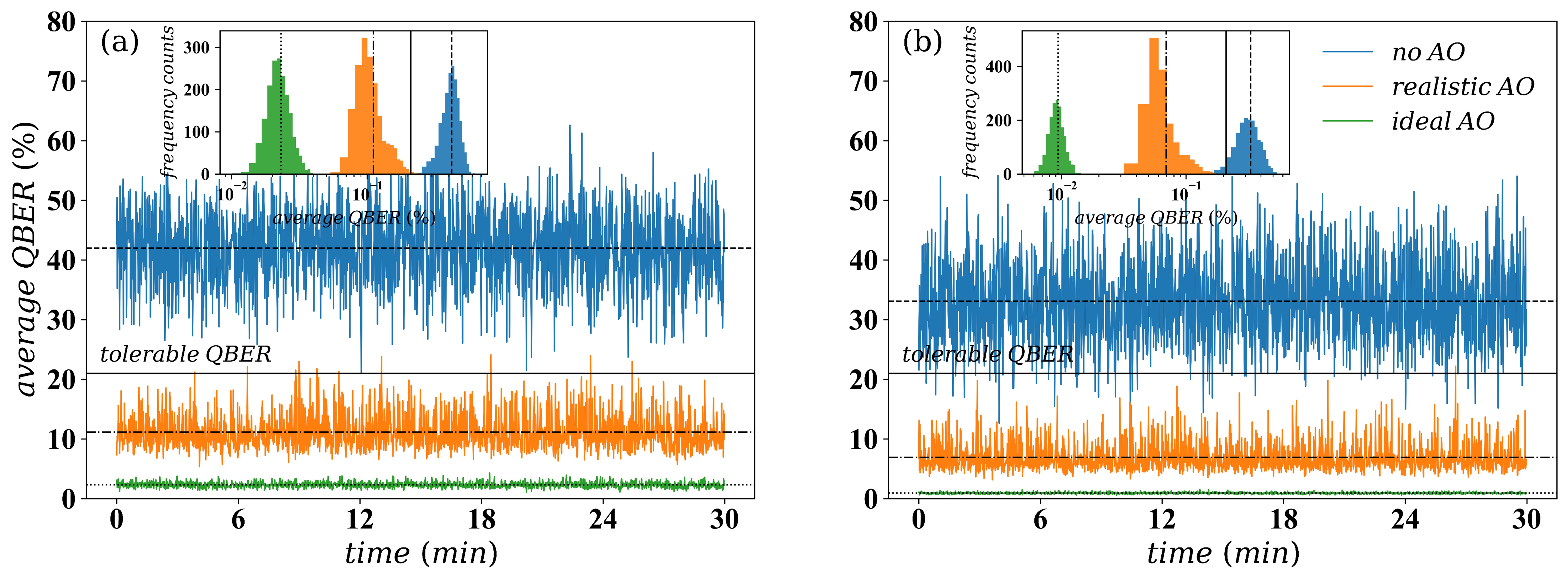}
	\caption{Average QBERs without, with realistic and with ideal AO correction
recorded during half an hour for $5$-dimensional OAM-based QKD system. Panel
(a) and (b) are realized without and with increasing mode spacing in
encoding subspace respectively. The inset is a QBER frequency histogram
counted from the variation curves of the average QBER. The dashed,
dash-dotted and dotted lines in (a), (b) denote the mean value of the
average QBER without ((a) $\overline{Q}=41.9\%$ (b) $\overline{Q}=33.1\%$),
with realistic ((a) $\overline{Q}=11.1\%$ (b) $\overline{Q}=6.9\%$) and with
ideal ((a) $\overline{Q}=2.3\%$ (b) $\overline{Q}=0.9\%$) correction
respectively, where $\overline{Q}\equiv \frac{1}{M}\sum_{i=1}^{M}Q^{\left(
i\right) }$, averaged over $M$ realizations of turbulence. The solid line
represents the tolerable QBER threshold of $5$-dimensional OAM-based QKD.
This simulation allows the correction capability of realistic AO to $30$%
-order and the turbulence level to $C_{n}^{2}=2.2\times 10^{-15}m^{-2/3}$.}
\end{figure}

\begin{figure}
	\centering
	\includegraphics[width=\linewidth]{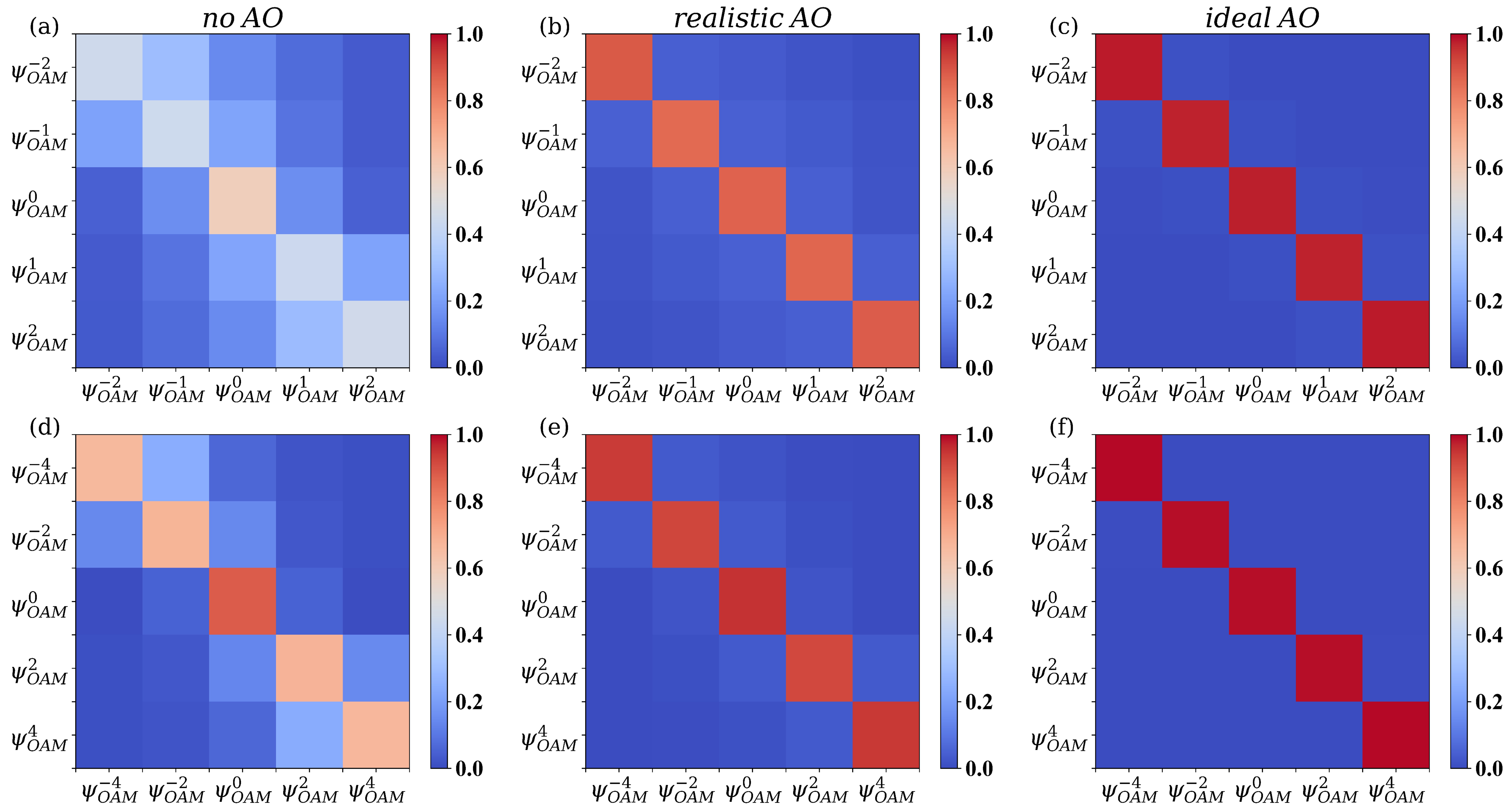}
	\caption{Measured crosstalk matrices of the OAM basis (a), (d) without, (b),
(e) with realistic, and (c), (f) with ideal correction. The upper and lower
three plots represent the realization without and with increasing mode
spacing respectively. All parameter settings are same as in Fig. 5.}
\end{figure}

As illustrated in Fig. 5(a), we find that, without increasing mode spacing
(i.e., using $\left\vert -2\right\rangle $, $\left\vert -1\right\rangle $, $%
\left\vert 0\right\rangle $, $\left\vert 1\right\rangle $ and $\left\vert
2\right\rangle $ for encoding), the average QBER can be mitigated from $%
41.9\%$ with a standard deviation of $6.14\%$ to $11.1\%$ with a standard
deviation of $2.93\%$ through the realistic AO correction. Notably, for the
ideal circumstance, the average QBER is reduced by $39.63\%$. With the help
of mode spacing increasing (i.e., using $\left\vert -4\right\rangle $, $%
\left\vert -2\right\rangle $, $\left\vert 0\right\rangle $, $\left\vert
2\right\rangle $ and $\left\vert 4\right\rangle $ for encoding, as shown in
Fig. 5(b)), we can alleviate the average QBER from $41.9\%$ to $33.1\%$
without correction, $11.1\%$ to $6.9\%$ with realistic correction and $2.3\%$
to $0.9\%$ with ideal correction, which indicates that the combination of
two strategies can lead to a significant reduction of average QBER caused by
atmospheric turbulence and rebuild a secure channel between two
communication parties. In Fig. 6, we also evaluate the measured crosstalk
matrices of perturbed OAM basis under the same settings of Fig. 5 without
and with AO correction respectively. More detailed discussions about the
crosstalk matrices are similar to the analysis presented in Fig. 5.

However, will AO correction work so well in the real-world? Unfortunately,
when we implement our enhanced AO-aided OAM-based QKD experiment, the
instrument inevitably generates some intrinsic noises (such as dark counts%
\cite{c29}, shot noise\cite{c73}, afterpulsing effect\cite{c87}, time delay
effect\cite{c54,c69,c70}, to name just a few); besides, the two MUBs will
also encounter different degrees of loss due to the receiving aperture\cite%
{c73}, atmospheric aerosols and dust particles\cite{c106} during the
propagation (the last two factors are not considered in this contribution,
see reasons in Ref. \cite{c107}). The accumulation of these realistic noises
can remarkably lead to an increased average QBER and seriously degrade the
performance of realistic AO system. Hence, we have to emphasize that the
reason why AO correction can achieve the high-quality enhancement in the
above simulation is that the above noise contributions are not considered,
which will be mentioned in subsection 5.

\subsection{Under arbitrary turbulence strengths}
In this section, we describe how the average QBER and the secret key rate of
OAM-based QKD change with respect to the atmospheric coherence length $r_{0}$
under different dimensions of encoding in Fig. 7, where each column from
left to right represents the results without (Fig. 7(a) and (d)), with
realistic (Fig. 7(b) and (e)) and with ideal (Fig. 7(c) and (f)) AO
correction respectively. The dashed vertical lines located at $r_{0}\approx
0.017m$ in all six plots correspond to the onset of weak scintillation. The
atmospheric and optical parameter settings adopted in this subsection are
listed as follows: $v=1m/s$, $\theta =\pi /2$, $z=1000m$, $\omega _{0}=0.03m$
and $\lambda =632nm$.

\begin{figure}
	\centering
	\includegraphics[width=\linewidth]{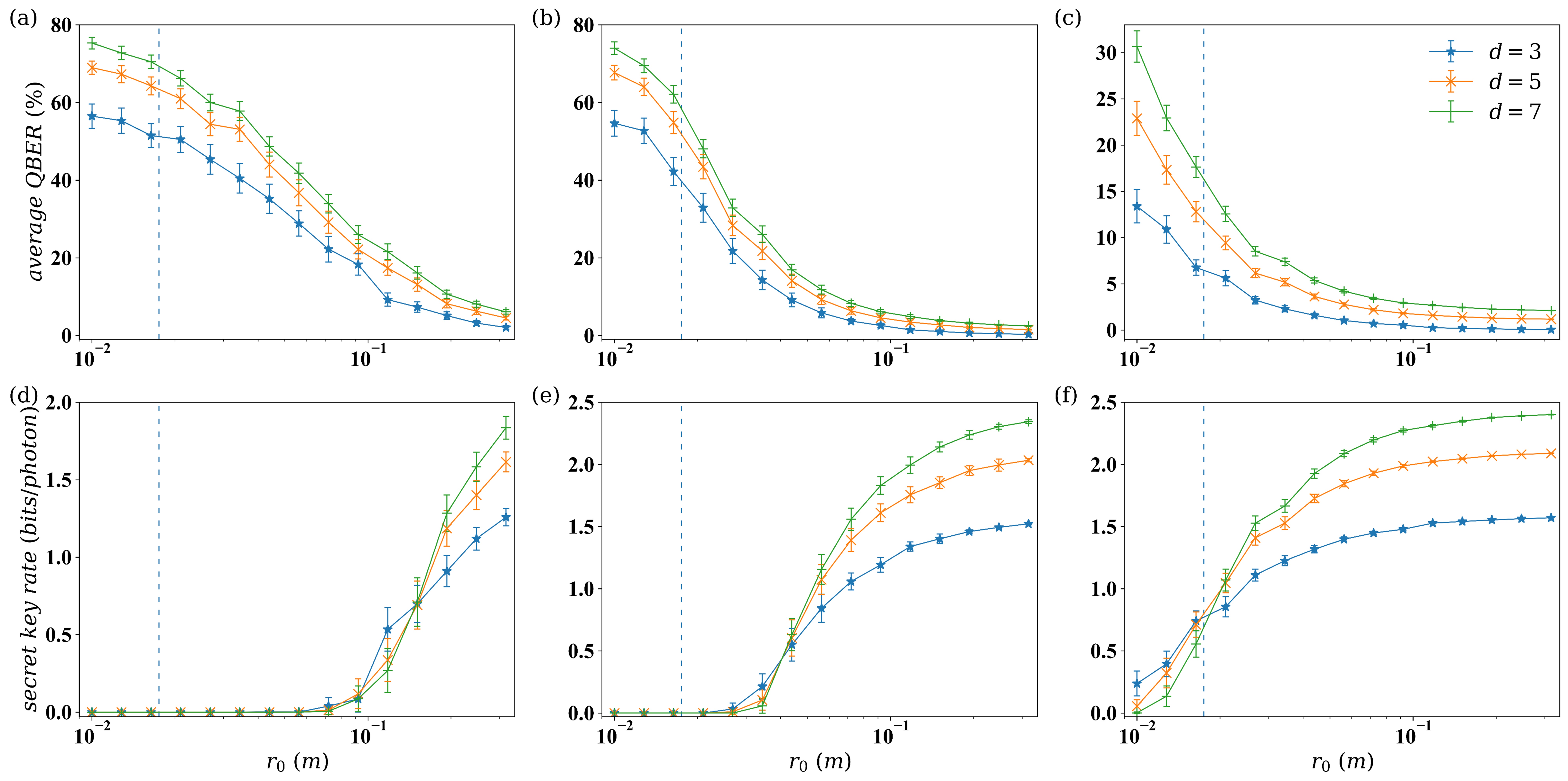}
	\caption{Average QBER (a)--(c) and secret key rate (d)--(f) of OAM-based QKD
as a function of $r_{0}$ with different dimensions of encoding. Different
degrees of correction are considered: (a), (d) no AO; (b), (e) realistic AO;
(c), (f) ideal AO, averaged over $300$ realizations of turbulence. The
dashed vertical lines at $r_{0}\approx 0.017m$ in all six plots correspond
to $\sigma _{R}^{2}=1$. The error bars represent the standard error. The
correction capability of AO and mode spacing is set to $30$-order and $1$ in
both cases.}
\end{figure}

From the results presented in Fig. 7(a), we observe that the average QBER
gradually decreases as the turbulence becomes weaker. For a higher
dimensional OAM-based QKD, the average QBER is overall superior to the lower
dimensional encoding in the strong scintillation regime, which is likely
explained by the fact that the OAM eigenstates with a higher azimuthal index
have a larger beam size so that they are more susceptible to atmospheric
turbulence. In the weak scintillation regime, all these differences between
different curves become smaller with decreasing turbulence strength.
Besides, as illustrated in Fig. 7(d), we see that no positive key rate can
be obtained by OAM-based QKD in all considered dimensions under moderate to
strong turbulence strengths (i.e., $r_{0}$ ranges from $0.01m$ to $0.06m$),
which causes both parties cannot build a secure communication link.
Generally, we expect to improve the system security and information capacity
of OAM-based QKD when we employ high dimensional encoding. However, we find
in Fig. 7(d) that when $r_{0}\approx 0.12m$, a lower secret key rate will be
obtained despite a higher dimensional encoding subspace is used (i.e., a
more secure communication link guaranteed by a higher dimensional OAM-based
QKD is destroyed by atmospheric turbulence). Fortunately, high dimensional
OAM-based QKD will reveal its unique advantages as the turbulence strength
gradually decreases. Therefore, it should be noted that, for a given
turbulence strength, selecting an appropriate dimension for encoding is
especially critical to acquire a better performance of OAM-based QKD (e.g.,
we choose $d=3$ for $r_{0}\approx 0.12m$ and $d=7$ for $r_{0}\approx 0.25m$).

\begin{figure}
	\centering
	\includegraphics[width=\linewidth]{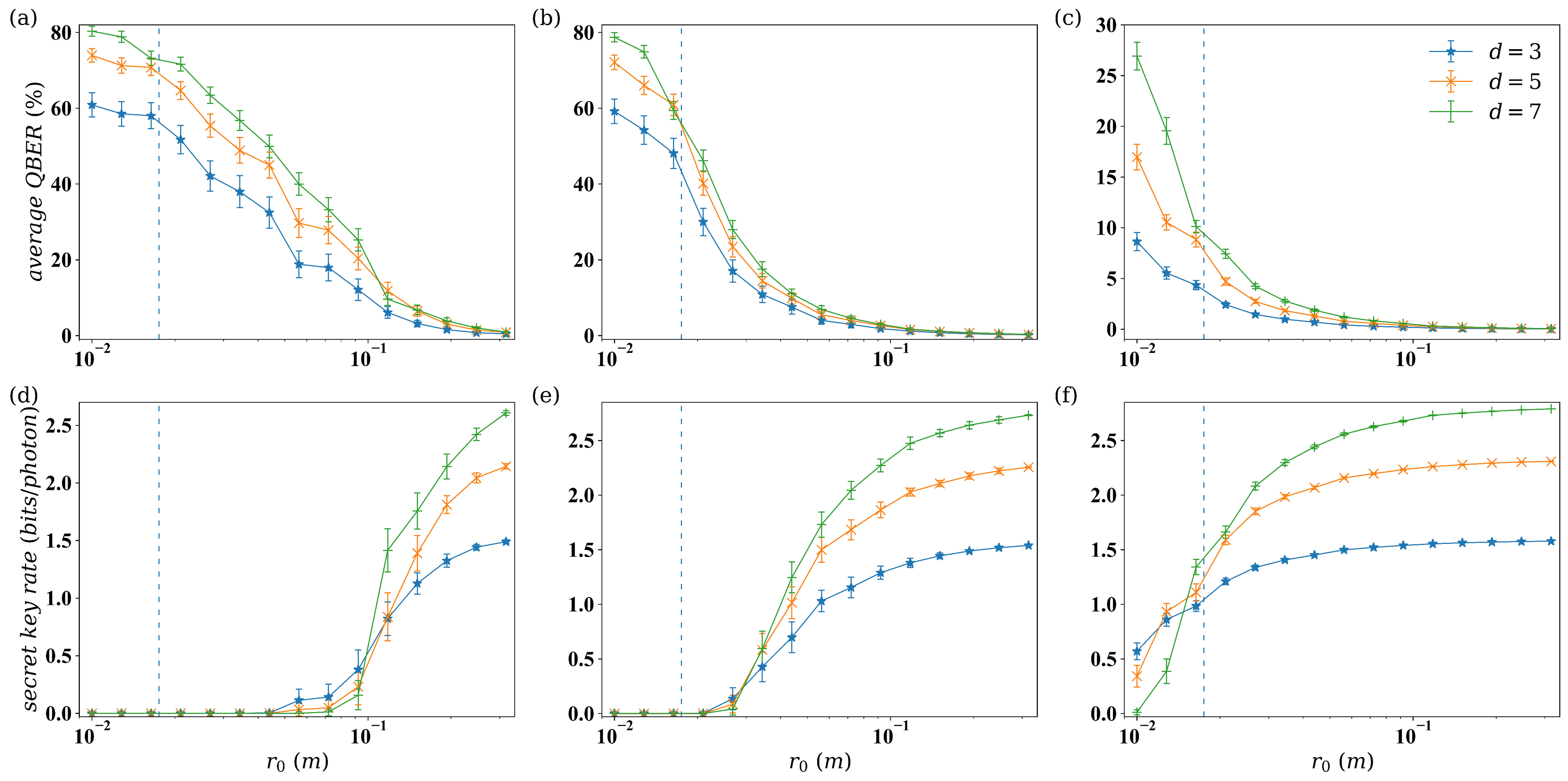}
	\caption{Average QBER (a)--(c) and secret key rate (d)--(f) of OAM-based QKD
as a function of $r_{0}$ with different dimensions of encoding. Different
degrees of correction are considered: (a), (d) no AO; (b), (e) realistic AO;
(c), (f) ideal AO, averaged over $300$ realizations of turbulence. The
dashed vertical lines at $r_{0}\approx 0.017m$ in all six plots correspond
to $\sigma _{R}^{2}=1$. The error bars represent the standard error. The
correction capability of AO and mode spacing is set to $30$-order and $2$ in
both cases.}
\end{figure}

The central and last columns in Fig. 7 illustrate the QKD's performance
achieved with realistic and ideal AO correction. Based on the comparison
between Fig. 7(a) and (b), we see that, in the weak scintillation regime,
the average QBER significantly decreases as the turbulence becomes weaker,
which implies that our enhanced AO is effective for improving the quality of
OAM-based QKD. Specifically, we notice that the realistic AO can achieve the
best performance enhancement under moderate turbulence strengths (e.g., when 
$r_{0}=0.044m$ and $d=5$, the average QBER decreases from $44\%$ to $14\%$
and the secret key improves from no positive keys to $0.605$ bits/photon).
On the other hand, we re-evaluate in Fig. 7(e) how the secret key rate
changes with respect to $r_{0}$ under the realistic AO correction. Compared
to Fig. 7(d), we observe that the realistic AO correction can lead to a
positive key rate when the turbulence strength is moderate, which
substantially improves the utilization of OAM-based QKD. Furthermore, with
the help of ideal correction, we obtain an upper boundary (Fig. 7(c) and
(f)) for the performance enhancement of OAM-based QKD, which leads to a
positive key rate in all considered turbulence strengths (e.g., when $%
r_{0}=0.02m$ and $d=5$, the secret key rate improves from no positive keys
to $1.04$ bits/photon).

We also enhance the performance of OAM-based QKD through increasing mode
spacing in encoding subspace. Comparing the results obtained between the two
strategies (Fig. 7 and 8), we note that, in the strong scintillation regime,
when the mode spacing becomes $2$, the average QBER of OAM-based QKD
counterintuitively increases compared to the successive encoding (see the
comparison between Fig. 7(a) and 8(a), e.g., when $r_{0}=0.01m$, the average
QBER without AO correction improves from $75.32\%$ to $80.32\%$ for $d=7$, $%
68.98\%$ to $73.95\%$ for $d=5$ and $56.49\%$ to $60.88\%$ for $d=3$, more
explanations are presented in Ref. \cite{c38}). However, such performance
degradation does not appear when we employ the ideal correction (see the
comparison between Fig. 7(c) and 8(c)), which is a compromise result between
the degradation and correction. Besides, in the weak scintillation regime,
we conclude that increasing the mode spacing can partially improve the
quality of OAM-based QKD, especially with the help of AO (e.g., when $%
r_{0}=0.15m$ and $d=5$, the secret key rate improves from $1.85$ bits/photon
to $2.1$ bits/photon with realistic AO correction and from $2.04$
bits/photon to $2.27$ bits/photon with ideal AO correction).

\subsection{Under arbitrary correction orders}
Since the correction capability of realistic AO system is commonly
determined by its order, we evaluate how the performance of an AO-aided
OAM-based QKD changes with respect to the correction order under different
dimensions of encoding in Fig. 9. We set $C_{n}^{2}=1.01\times
10^{-14}m^{-2/3}$ during the simulation. It can be seen that the quality of
OAM-based QKD improves significantly as the correction order of AO
increases. However, such significant performance enhancement is only in the
lower order correction range. The primary reason because the
turbulence-induced aberrations is almost entirely concentrated in the
low-frequency part. On the other hand, since we set the turbulence strength
in the moderate to strong range, we notice that when the correction order is
larger than $100$-order, the quality of OAM-based QKD still improves
somewhat as the correction order increases. Generally, we anticipate that
the quality of OAM-based QKD might not be significantly enhanced with the
increase of the correction order of AO under the weak turbulence strengths.
In Fig. 9(b) and 9(d), we also present the results of increasing mode
spacing in encoding subspace. We observe that only a minor performance
enhancement appears compared to the successive encoding.

\begin{figure}[h]
	\centering
	\includegraphics[width=0.85\linewidth]{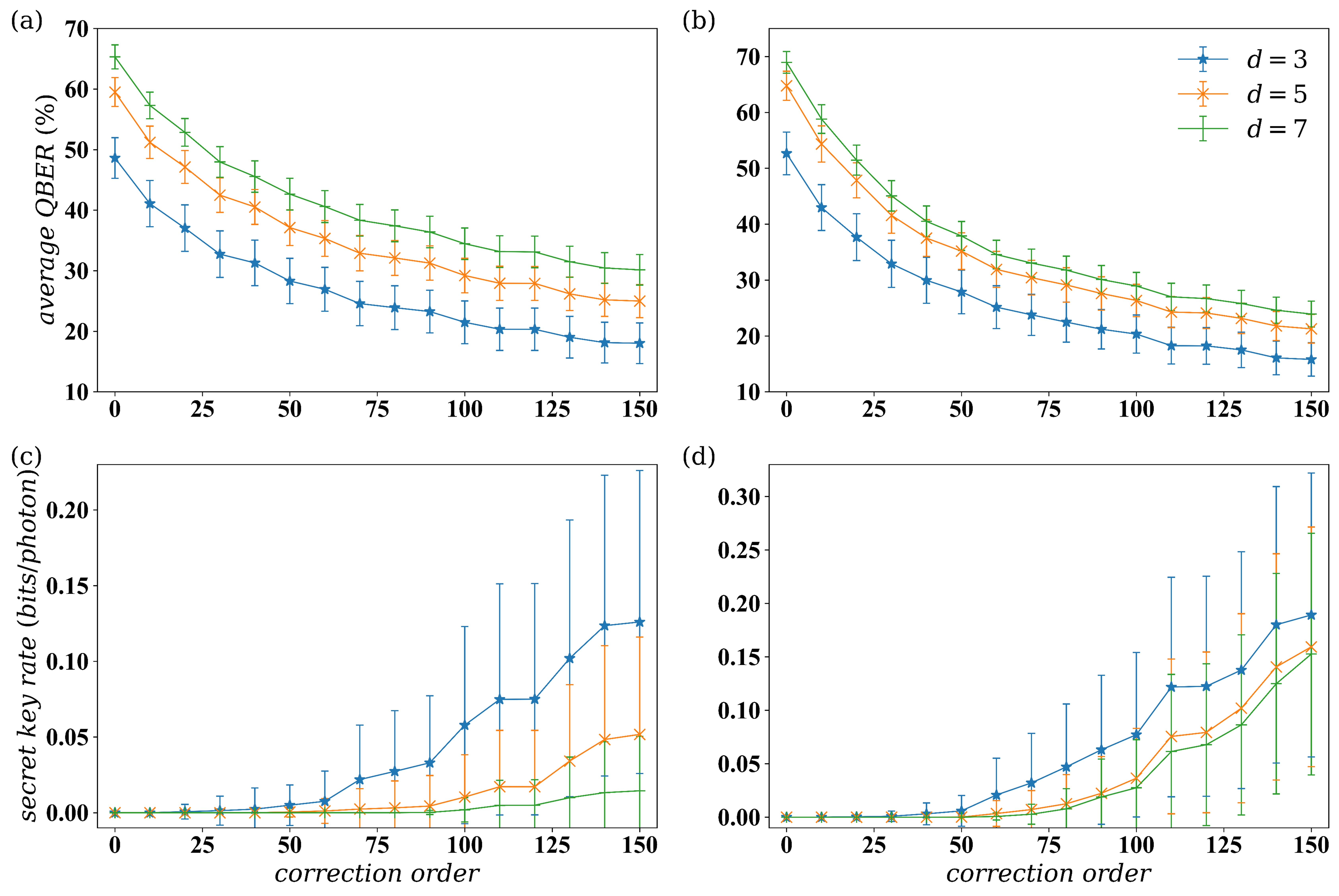}
	\caption{Average QBER (a), (b) and secret key rate (c), (d) of OAM-based QKD
as a function of correction order with different dimensions of encoding.
Panel (a), (c) and panel (b), (d) represent the results realized without and
with increasing mode spacing respectively. All calculations in four plots
are 300 realizations of turbulence. The error bars represent the standard
error. The turbulence level in four plots is $C_{n}^{2}=1.01\times
10^{-14}m^{-2/3}$.}
\end{figure}

Another feature can be seen in Fig. 9(c) and (d) is that when we increase
the correction order of AO and mode spacing in encoding subspace, we can
recover the advantage of high-dimensional encoding to some extent, but
choosing $3$-dimensional OAM states for encoding is still the best choice
under this turbulence strength.

\subsection{Realistic considerations}

Now, we evaluate the performance enhancement achieved by our enhanced AO
with more realistic considerations, including the effects of loss from a
limited sized receiving aperture and time delay between wavefront
measurement and correction.

\subsubsection{Loss}

The state-dependent loss\cite{c73,c109} caused by the diffraction effects
significantly degrades the performance of OAM-based QKD in long-distance
propagation. In the two MUBs, an OAM eigenstate with a larger azimuthal
number suffers from more significant loss and acquires more propagation phase%
\cite{c83} due to the limited size of receiving aperture, leading to a
reduced quality of QKD even in the absence of turbulence. In order to
evaluate the impact of state-dependent loss on the performance of OAM-based
QKD, we re-examine how the average QBER and the secret key rate change with
respect to $r_{0}$ under different sizes of receiving aperture. Here, we
only consider the performance degradation in 5-dimensional OAM-based QKD.
the other parameter settings adopted during the simulation is the same as
that in subsection 3.

\begin{figure}
	\centering
	\includegraphics[width=\linewidth]{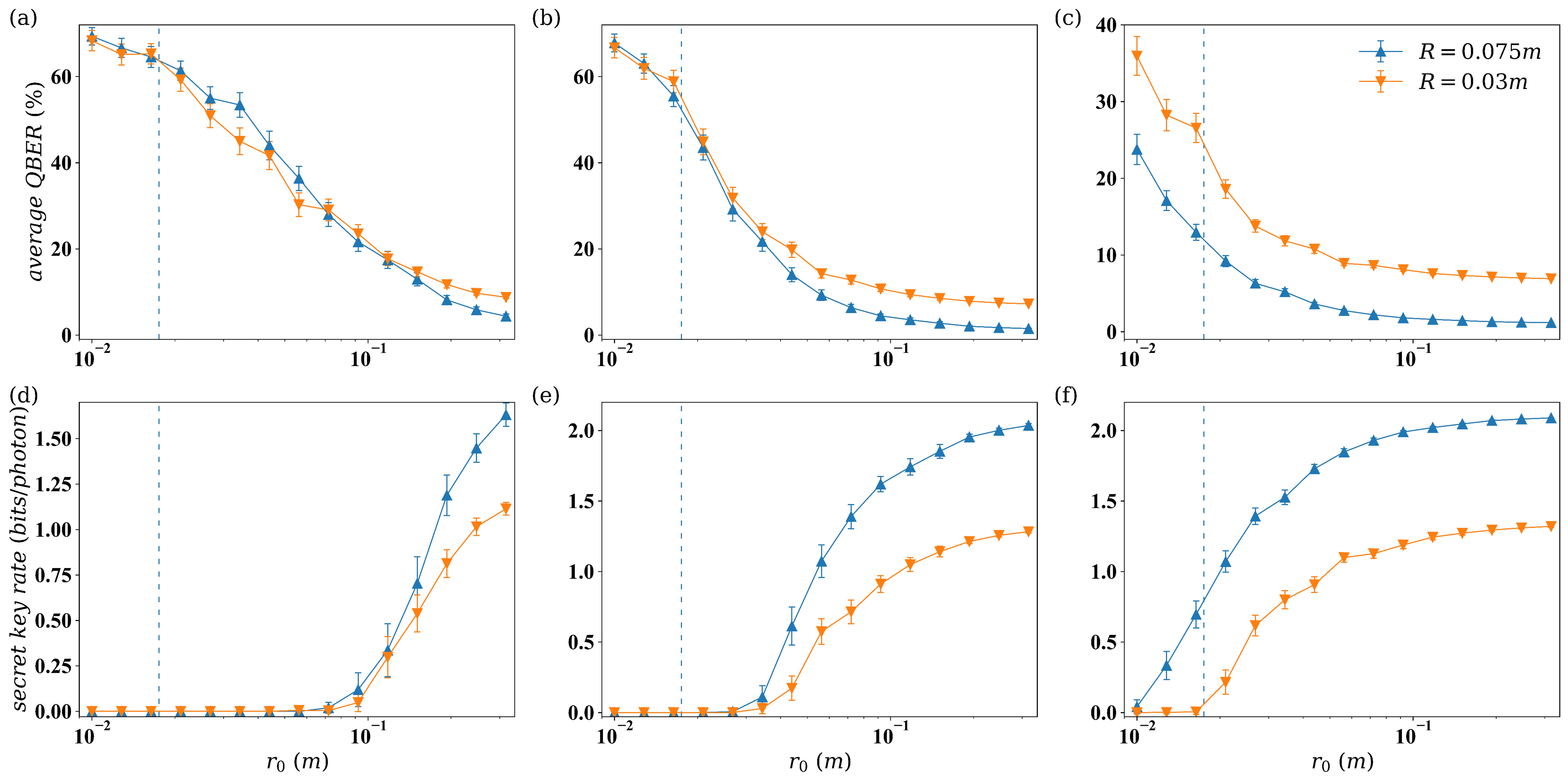}
	\caption{Average QBER (a)--(c) and secret key rate (d)--(f) of $5$%
-dimensional OAM-based QKD as a function of $r_{0}$ with different sizes of
receiving aperture. Different degrees of correction are considered: (a), (d)
no AO; (b), (e) realistic AO; (c), (f) ideal AO, averaged over $300$
realizations of turbulence. The dashed vertical lines at $r_{0}\approx
0.017m $ in all six plots correspond to $\sigma _{R}^{2}=1$. The error bars
represent the standard error. All other parameter settings are same as in
Fig. 7.}
\end{figure}

As illustrated in Fig. 10(a), we observe that when the channel is under
strong turbulence strengths and without AO correction, reducing the
receiving aperture can surprisingly mitigate the impact of atmospheric
turbulence to some extent (e.g., when $r_{0}=0.034m$, the average QBER
decreases from $53.4\%$ to $44.98\%$), which is somewhat counterintuitive.
The main reason may be that, under these turbulence strengths, the smaller
size of the aperture is similar to a low-pass filter\cite{c110}, a large
amount of turbulence-induced vortex-antivortex pairs may wander outside the
receiving aperture and result in a decrease of average QBER\cite{c111,c112}.
Conversely, in the presence of AO correction, we notice that decreasing the
size of receiving aperture will lead to a significant increase to average
QBER (Fig. 10(b) and (c), e.g., when $r_{0}=0.01m$, the average QBER
increases from $23.77\%$ to $35.93\%$ with ideal AO correction) and decrease
to secret key rate (Fig. 10(e) and (f), e.g., when $r_{0}=0.31m$, the secret
key rate decreases from $2.89$ bits/photon to $1.32$ bits/photon with ideal
AO correction). Such an observation is likely because the aperture loss
becomes the primary source of bit error for OAM-based QKD after implementing
AO correction.

\subsubsection{Time delay}

In above simulation, we investigate the quality enhancement of OAM-based QKD
based on the assumption that AO correction is implemented in real time.
However, since our enhanced AO includes the procedure of reconstructing and
unwrapping the phase of beacon light, the impact of time delay effect
between wavefront measurement and correction on OAM-based QKD cannot be
ignored, which can also significantly affect the performance of OAM-based
QKD. It is well-known that real-time implementation can not be realized
except the Greenwood frequency is small enough for the bandwidth of AO system%
\cite{c113}. In order to re-evaluate the impact of performance degradation
caused by the time delay effect on OAM-based QKD, we set the bandwidth of AO
as one-fifth of the sampling frequency of a Hartmann camera\cite{c114}
(i.e., if we assume the camera sample frequency equals to $1KHz$, then $%
f_{AO}=200Hz$), and change the Greenwood frequency $f_{G}$ by adjusting the
wind velocity through the following relationship\cite{c54}%
\begin{equation}
f_{G}=\frac{0.43v}{r_{0}}.  \label{eq15}
\end{equation}

\begin{figure}
	\centering
	\includegraphics[width=0.75\linewidth]{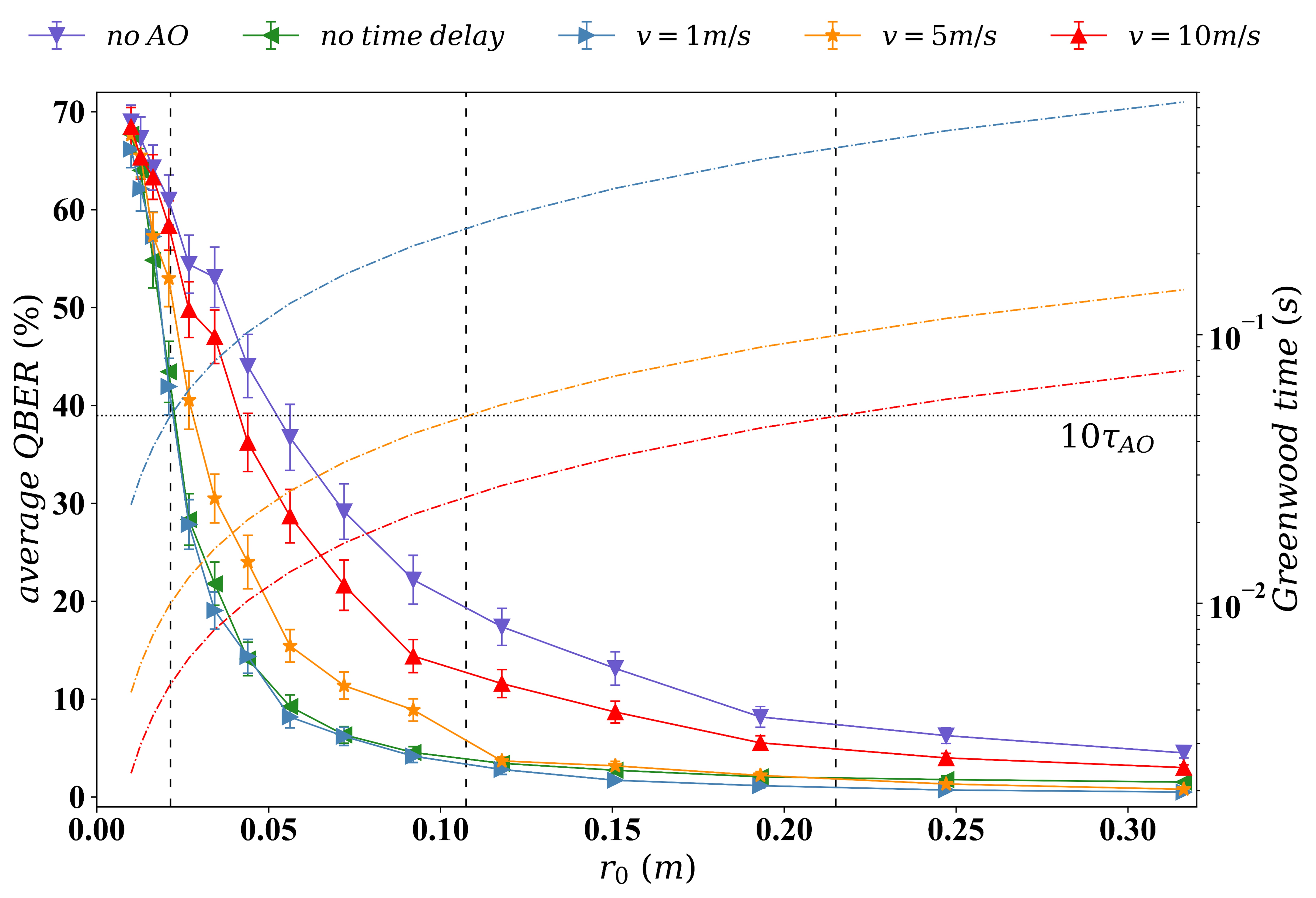}
	\caption{Average QBER of $5$-dimensional OAM-based QKD as a function of $%
r_{0}$ with different wind velocities, averaged over $300$ realizations of
turbulence. The horizontally dotted line represents the ten-times time delay
of AO (defined as $\tau _{AO}\equiv 1/f_{AO}$) if we assume the bandwidth of
AO is $200Hz$ (i.e., $f_{AO}=200Hz$). The dash-dotted lines denote the
Greenwood time $\tau _{G}$ according to Eq. (\ref{eq15}), Refs. \cite%
{c38,c58} indicate that the crossing point between the dash-dotted lines and
the dotted line represents the bandwidth of AO is large enough to provide a
comprehensive correction. The dashed lines represent the value of $r_{0}$
corresponding to the intersection of $\tau _{G}$ and $10\tau _{AO}$ for
different wind velocities. The correction capability of AO is same as in
Fig. 7. The error bars represent the standard error.}
\end{figure}

In Fig. 11, we consider how the average QBER changes with respect to $r_{0}$
under different wind velocities. We observe that when the wind velocity is
small (i.e., $v=1m/s$), the evolution of average QBER almost coincides with
the results obtained without time delay. On the contrary, with the increase
of wind velocity, the average QBER starts to deviate from the no time delay
counterparts, leading to the performance degradation of OAM-based QKD (e.g.,
when $r_{0}=0.026m$, the average QBER increases from $28.34\%$ to $40.54\%$
for $v=5m/s$, from $28.34\%$ to $49.78\%$ for $v=10m/s$). Besides, we also
notice that when the turbulence becomes weaker, the deviation together
inclines to disappear. To explain above phenomena, Fig. 11 is horizontally
divided into two parts\cite{c115}. We observe that when $v=5m/s$ and the
Greenwood time (defined as $\tau _{G}\equiv 1/f_{G}$) passes this threshold,
the bandwidth of AO is large enough to provide a comprehensive correction
(see more discussions in Ref. \cite{c38}). However, these conclusions are
not established for $v=10m/s$, which implies that, in the large wind
velocity regime, the bandwidth of AO needs to be enhanced to satisfy the
fast time-varying wavefront sampling and compensation. Finally, by comparing
to the results obtained without AO correction, we see that AO remains able
to mitigate the performance degradation even though it works in the large
wind velocity environment.

\section{Discussion and conclusion}

In this paper, we employ the ILPS method to simulate the performance
degradation of OAM-based QKD under arbitrary turbulence strengths. The main
idea behind ILPS is to use the partial data from previous phase screen to
generate the results of next one, which builds a connection between two
generated phase screens. An advantage of this design is that the wind
velocity of turbulence can be simulated by adjusting the amount of motion in
each iteration, which provides a feasible way to simulate the dynamical
process of incident light propagation and how AO dynamically corrects the
turbulence-induced aberrations in real time. However, we have to highlight
the shortcoming of ILPS, which uses the spatial correlation of phase screen
between two iterations to replace the temporal correlation of turbulence
variation (e.g., the variation of wind velocity increases the Greenwood
frequency of turbulence\cite{c54,c70}, however, ILPS does not perform well
in the face of turbulence variation in a windless environment), a topic
needs to be addressed in the future.

Considering the inability of traditional AO while encountering phase cuts%
\cite{c66,c67,c68}, we demonstrate the feasibility of AO that includes the
wrapped cuts elimination, which significantly alleviates the impact of
atmospheric turbulence on OAM-based QKD. We show that, in the weak
scintillation regime, the employment of realistic AO for correcting the
turbulence-induced aberrations can lead to a $30\%$ performance enhancement
approximately. However, in the strong scintillation regime, we observe a
rapid reduction of the performance of AO correction. Furthermore, we suggest
that if we wish to obtain a high-quality performance enhancement, an
advanced AO system is necessary for the realistic experiments\cite{c38}.

Since the additional noise contributions (such as state-dependent loss, time
delay effect of AO, to name just a few) degrade the high-quality enhancement
achieved by AO, we re-evaluate the correction effect of AO and conclude that
AO can still rebuild a secure communication channel and recover the high
information capacity of OAM-based QKD, even in the increased loss and wind
velocity environment. However, for the real-time correction consideration,
we still need to enhance the bandwidth of AO when the wind velocity becomes
larger.

Finally, it should be noted that although our enhanced AO includes the
procedure for eliminating wrapped cuts, the branch cuts caused by
turbulence-induced aberrations cannot be entirely negated by traditional AO
techniques. Further research on minimizing the length of branch cuts that
occur in the perturbed phase may prove fruitful.\\

\section*{Funding} Anhui Provincial Natural Science Foundation (1908085QA37); National Natural Science Foundation of China (11904369); State Key Laboratory of Pulsed Power Laser Technology Supported by Open Research Fund of State Key Laboratory of Pulsed Power Laser Technology (2019ZR07).\\

\section*{Acknowledgments} The authors are very grateful to the reviewers for their valuable comments. Z. T. thanks Prof. Ruizhong Rao for his careful reading and insightful suggestions.\\

\section*{Disclosures} The authors declare no conflicts of interest.\\
%%%%%%%%%% If using BibTeX:

%%%%%%%%%% If preparing manually:
% \newcommand{\enquote}[1]{``#1''}

% \bibitem{Zhang:14}
% Y.~Zhang, S.~Qiao, L.~Sun, Q.~W. Shi, W.~Huang, L.~Li, and Z.~Yang,
%   \enquote{Photoinduced active terahertz metamaterials with nanostructured
%   vanadium dioxide film deposited by sol-gel method,}
%   {\protect\JournalTitle{Optics Express}} \textbf{22}, 11070--11078 (2014).

% {Optical Society}, \enquote{{OSA Publishing},}
%   \url{http://www.osapublishing.org}.

% \bibitem{FORSTER2007}
% P.~Forster, V.~Ramaswamy, P.~Artaxo, T.~Bernsten, R.~Betts, D.~Fahey,
%   J.~Haywood, J.~Lean, D.~Lowe, G.~Myhre, J.~Nganga, R.~Prinn, G.~Raga,
%   M.~Schulz, and R.~V. Dorland, \enquote{Changes in atmospheric consituents and
%   in radiative forcing,} in \enquote{Climate Change 2007: The Physical Science
%   Basis. Contribution of Working Group 1 to the Fourth assesment report of
%   Intergovernmental Panel on Climate Change,}  S.~Solomon, D.~Qin, M.~Manning,
%   Z.~Chen, M.~Marquis, K.~B. Averyt, M.~Tignor, and H.~L. Miler, eds.
%   (Cambridge University Press, 2007).

% \end{thebibliography}


\begin{thebibliography}{999}
\bibitem{c71} It is worth highlighting that there is no essential difference
between a conventional Hartmann wavefront sensor and a Shack-Hartmann
wavefront sensor. The latter only replaces the pinhole of a conventional
wavefront sensor with a microlens array (i.e., the former is
pinhole-imaging, while the latter is convex-lens-imaging).
\bibitem{c72} It is noted that LG modes chosen here for encoding is not the
best choice, but a compromise consideration. Actually, in order to reduce
the loss due to mode dependent diffraction, we commonly choose the
eigenstate of $H$ (i.e., the mode with the lowest loss for free-space
propagation, for more details about the eigenstate and definition of $H$, we
refer the reader to Refs. \cite{c73,c74}) as a probe state for OAM encoding%
\cite{c75}, which is consistent with the mathematical solution of the
communication modes proposed in \cite{c76} under the circular aperture. When
the Fresnel number becomes larger, the eigenstates of $H$ degenerate to LG
modes. Moreover, the communication implemented by LG modes can achieve a
comparable performance compared to the eigenstates of $H$ under weak
turbulence strengths\cite{c77}.
\bibitem{c86} Unlike other works, here, we also consider the radial-mode
scrambling caused by atmospheric turbulence. Since the information is
encoded by the photonic OAM state, we restrict our attention to the
crosstalk probability within the OAM subspace.
\bibitem{c92} It is noted that the validity of ILPS is guaranteed by
calculating its phase structure function. As discussed in Ref. \cite{c63}, a
great coincidence between the average structure function and theoretical one
is commonly obtained by setting $N_{col}\geq 2$.
\bibitem{c94} It is worth noting that the effect of atmospheric turbulence
on beam propagation is wavelength dependent. We assumed the wavelength of
beacon light is close enough to signal modes in this paper. Concretely, a
beacon light whose wavelength around $100nm$ shorter than signal modes is
acceptable\cite{c54,c56,c58}.
\bibitem{c101} The realistic AO correction mentioned in this paper is the
results after eliminating the wrapped cuts.
\bibitem{c102} For the weak and strong scintillation regime, we set $%
C_{n}^{2}=2.2\times 10^{-15}m^{-2/3}$ and $3.22\times 10^{-14}m^{-2/3}$,
which is equivalent to $r_{0}=0.05m$ and $r_{0}=0.01m$ when $z=1000m$. It is
noted that $r_{0}$ is calculated with a fixed wavelength of $500nm$, which
is not same as the intrinsic wavelength of the incident states. The reason
why we chose $500nm$ for the calculation is that $r_{0}$ is associated with
wavelength according to the Eq. (3) in Ref. \cite{c51}. In atmospheric optics, a wavelength of $500nm$ is generally specified to facilitate the use of $r_{0}$
to measure the turbulence strength of a communication link.
\bibitem{c105} It is highlighted that, in a realistic AO system, the
correction order is closely associated with the number of actuator onto a
deformable mirror. We set the correction capability of AO to $30$-order for
the realistic consideration.
\bibitem{c107} The impact of aerosol-particle scattering on the performance
of OAM-based QKD can be described by adjusting the size of receiving
aperture. An interesting topic is that whether eavesdroppers can steal
information through scattered light (see a similar discussion in Ref. \cite%
{c108}).
\bibitem{c114} It is noted that we generally set the bandwidth of AO as $2$--%
$3$ times the sample period of a Hartmann camera. Here we set to $5$ because
we consider the time delay caused by phase unwrapping.
\bibitem{c115} Refs. \cite{c38,c58} indicate that a comprehensive correction
can be obtained if the Greenwood time of turbulence is ten times larger than
the time delay of AO.
\end{thebibliography}
\end{document}